# Possible emergence of a skyrmion phase in ferroelectric GaMo$_4$S$_8$

Hui-Min Zhang,[1, 2] Jun Chen,[1] Paolo Barone,[2] Kunihiko Yamauchi,[3] Shuai Dong,[1,*] Silvia Picozzi[2,†]

[1] *School of physics, Southeast University, Najing 211189, China*

[2] *Consiglio Nazionale delle Ricerche, Institute for Superconducting and Innovative Materials and devices (CNR-SPIN), c/o Univ. "G. D'Annunzio", Chieti 66100, Italy*

[3] *Institute of Scientific and Industrial Research, Osaka University, Ibaraki, Osaka 567-0047, Japan*

**Abstract**

Polar lacunar spinels, such as GaV$_4$S$_8$ and GaV$_4$Se$_8$, were proposed to host skyrmion phases under magnetic field. In this work, we put forward, as a candidate for Néel-type skyrmion lattice, the isostructural GaMo$_4$S$_8$, here systematically studied via both first-principles calculations and Monte Carlo simulations of model Hamiltonian. Electric polarization, driven by Jahn-Teller distortion, is predicted to arise in GaMo$_4$S$_8$, showing a comparable size but an opposite sign with respect to that evaluated in V-based counterparts and explained in terms of different electron counting arguments and resulting distortions. Interestingly, a larger spin-orbit coupling of 4$d$ orbitals with respect to 3$d$ orbitals in vanadium-spinels leads to stronger Dzyaloshinskii-Moriya interactions, which are beneficial to stabilize a cycloidal spin texture, as well as smaller-sized skyrmions (radius<10 nm). Furthermore, the possibly large exchange anisotropy of GaMo$_4$S$_8$ may lead to a ferroelectric-ferromagnetic ground state, as an alternative to the ferroelectric-skyrmionic one, calling for further experimental verification.

\* Email: sdong@seu.edu.cn    † Email: silvia.picozzi@spin.cnr.it

I. **Introduction**

Magnetic skyrmions, showing whirl-like spin textures with nontrivial topology, have drawn increasing interests in recent years, due to their appealing potential applications in spintronic devices with low energy consumption [1]. The skyrmion lattice was mostly observed upon application of magnetic fields in metallic alloys with chiral structures, such as FeGe [2], MnGe [3], MnSi [4]. They all share the B20-type crystal structure (space group $P2_13$) [1], a weak magnetocrystalline anisotropy, and common phase diagrams with a helimagnetic ground state. Besides metallic alloys, skyrmions were also found in insulating oxides, such $Cu_2OSeO_3$, whose space group is $P2_13$ [5-7]. A magnetoelectric coupling is expected in insulating compounds hosting skyrmions, opening up the possibility to manipulate the latter by means of electric field [8].

Magnetic skyrmions can be classified as Bloch-type and Néel-type. When moving from the skyrmion core to the periphery, the spins rotate in a tangential plane (i.e. perpendicular to the radial direction) for Bloch-type skyrmions, whereas they rotate in a radial plane for Néel skyrmions. For the aforementioned chiral magnets, the skyrmions are Bloch-type, which have been extensively investigated. In contrast, the Néel-type skyrmions were predicted to exist in polar materials with the $C_{nv}$ symmetry and then found first in $GaV_4S_8$ and later in $GaV_4Se_8$ [9-14]. $GaV_4S_8$ and $GaV_4Se_8$ belong to the lacunar spinel $AM_4X_8$ family ($A$ = Ga and Ge; $M$ = V, Mo; $X$ = S and Se), whose low-temperature structure is rhombohedral (space group $R3m$) with $C_{3v}$ symmetry [15]. In early works, $GaV_4S_8$ was reported to be a ferromagnet [15-17], but later Nakamura *et al*. showed that $GaV_4S_8$ is not a simple ferromagnet below $T_C$ since antiferromagnetic signatures (ascribed to spin flips) appears above 5 K [18]. In recent studies, based on atomic force microscopy imaging, complemented with small-angle neutron scattering measurements and theoretical calculations, three magnetic phases were identified in $GaV_4S_8$: a cycloidal phase, a Néel-type skyrmion lattice phase, and a ferromagnetic phase [12,14,19]. The skyrmion lattice phase in $GaV_4S_8$ only exists in a very narrow parameter space: temperature ~10-13 K under ~50 mT magnetic field, as reported in Refs. [12,14]. On the other hand, the ground state of $GaV_4Se_8$ has been reported to display a cycloidal spin phase, implying that the skyrmion lattice can be stabilized in a broader temperature region down to 0 K [13].

Despite being also a member of the lacunar spinel $AM_4X_8$ family, $GaMo_4S_8$ has been the focus of fewer experimental studies. The very early works claimed it to be ferromagnetic below 19.5 K [20] and a recent theoretical calculation also mentioned ferromagnetism [21]. However, all these studies were done before the discovery of skyrmions in $GaV_4S_8$, the latter also been erroneously believed to be a simple ferromagnet [12,14]. Therefore, it seems necessary and timely to perform a systematic study, in order to recheck the physical properties of $GaMo_4S_8$. An appealing characteristic of $GaMo_4S_8$, as compared to V-based counterparts, is that the 4$d$ orbitals of Mo display larger spin-orbit coupling and more extended spatial distribution than 3$d$ orbitals of V, which may result in tuning the subtle balance between the ferromagnetic, cycloid, and skyrmion phases.

Figure 1(a) shows the minimal unit cell, containing weakly linked molecular units - cubane $(Mo_4S_4)^{5+}$, formed by two interpenetrating $Mo_4$ and $S_4$ tetrahedra [see Fig. 1(e)], and tetrahedral $(GaS_4)^{5-}$ as building blocks. According to early experiments, when decreasing the temperature, a structural transition from cubic (*F*-43*m*, No. 216) to rhombohedral structure (*R*3*m*, No. 160) occurs at 45.5 K in Ref. [22] and 47 K in Ref. [23], driven by the cooperative Jahn-Teller distortion. At high temperatures, the structure is noncentrosymmetric and nonpolar with the $T_d$ point group, i.e. the Mo-tetrahedron is regular with equivalent Mo ions and Mo-Mo bonds. In the low-temperature polar structure, the Mo-tetrahedron is compressed along the [111] direction, distinguishing the apical ($Mo_1$) ion from the planar ($Mo_2$) ions and lowering the symmetry to the $C_{3v}$ point group. Similar Jahn-Teller distortions occur in $GaV_4S_8$ and $GaV_4Se_8$ at 38 K [15] and 41 K [13] respectively, but their V-tetrahedra are elongated along the [111] direction, as opposed to the Mo-tetrahedra compression.

## II. **Methods**

Although Ref. [21] briefly mentioned a preliminary density functional theory (DFT) study of $GaMo_4S_8$, a comprehensive investigation is still lacking, especially regarding the possible skyrmion phase. Therefore, in this work, $GaMo_4S_8$ is systematically studied using both DFT and Monte Carlo (MC) simulations based on model Hamiltonian, in order to clarify the similarities and differences between Mo-based and V-based lacunar spinel materials.

The DFT method can deal with electronic structure as well as basic magnetic properties, in small cells (typically with dozens of ions) at zero temperature. However, it can not directly handle the skyrmions at finite temperature and under magnetic fields. Therefore, the model study is necessary, with physical coefficients extracted from the DFT calculations. Generally, MC method can be a powerful tool to determine the thermodynamics phases for a complicated magnetic system. By combining this two methods, it allows us to obtain more comprehensive physics of $GaMo_4S_8$, especially its magnetic properties.

*II.A DFT calculation*

First-principles DFT calculations were performed using the Vienna *ab initio* Simulation Package (VASP) [24,25] based on the projected augmented wave pseudopotentials. For the exchange-correlation functional, the PBEsol (Perdew-Burke-Ernzerhofrevised for solids) [26] parametrization of the generalized gradient approximation plus *U* (GGA+*U*) method [27-29] was used. The Hubbard $U_{eff}$ (=*U*-$J_H$, $J_H$: the Hund exchange) was imposed on Mo's 4*d* orbitals using the Dudarev approach [30] for the structural relaxation, polarization, electronic structure, as well as magnetic ground state without spin-orbit coupling (SOC). We also adopted the Liechtenstein approach, treating separately the *U* and $J_H$ parameters, [31] to calculate the magnetocrystalline anisotropic energy (MAE) including the anisotropic spin exchange (ASE) interactions and single ion anisotropy (SIA), as well as Dzyaloshinskii-Moriya (DM) interactions with SOC enabled.

The atomic positions and lattice constants were fully optimized until the Hellman-Feynman forces converged to less than 0.01 eV/Å. The plane-wave cutoff was set to 500 eV. The Monkhorst-Pack $k$-point meshes were chosen as 7×7×7 in the primitive rhombohedral structure and 6×6×3 in the hexagonal structure. Larger supercells, following the same procedure described in Ref. [32], were used to extract magnetic coefficients. The polarization was calculated by means of both the Berry phase method [33] and Wannier center displacements [34]. The WANNIER90 package was used to obtain the Wannier orbitals of the valence bands [35].

*II.B Monte Carlo simulation*

The Markov-chain MC method with Metropolis algorithm was employed to simulate the magnetic phase diagram. The MC simulation was done on a $N=36\times36\times3$ hexagonal close-packed lattice with periodic boundary conditions, which was large enough to recover the cycloid phase as well as skyrmions. Larger lattices (e.g. $N=60\times60\times3$) were also verified at some temperature and magnetic field points, confirming the physical results obtained for smaller lattices. For each simulation point, the initial $1\times10^5$ MC steps were discarded for equilibrium consideration and another $1\times10^5$ MC steps were retained for statistical averaging of the simulation. The quenching process [36] was used for temperature and magnetic field scanning.

To characterize the magnetic phase transitions, the specific heat $C$ and magnetization $M$ were calculated to determine the critical temperatures [37], which is defined as :

$$C = (<H^2> - <H>^2)/Nk_BT^2 \tag{1}$$

where $H$ is the thermal energy; $N$ is the number of spins; $k_B$ is Boltzmann constant, and $T$ is temperature. For additional insights on the spiral order and skyrmions, the spin structure factors were also calculated. Furthermore, to define the (in-plane) density of skyrmions, the local chirality $\chi_i$ was calculated as [4]:

$$\chi_i = [S_i \cdot (S_{i+\hat{x}} \times S_{i+\hat{y}}) + S_i \cdot S_{i-\hat{x}} \times S_{i-\hat{y}})]/8\pi \tag{2}$$

In the continuum limit, a single skyrmion gives $\chi = \sum_i \chi_i = 1$.

### III. **Results and discussion**

*III.A Structural and ferroelectric properties*

As a first step, it is necessary to check the basic physical properties of GaMo$_4$S$_8$. For simplicity, the DFT structural relaxation was performed imposing a ferromagnetic spin configuration. The relaxed lattice constant and rhombohedral angle are shown in Figs. 1(b) and 1(c), respectively, as a function of $U_{eff}$. These lattice parameters do not show a dramatic dependence on the Hubbard parameter, strengthening the reliability and solidity of our following results. As a reference, in a previous DFT study on GaV$_4$S$_8$ an effective Hubbard $U$-$J_H$=2 eV was applied on V's 3$d$ orbitals [21]. Considering the less localized character of 4$d$ orbitals of Mo, as opposed to vanadium 3$d$ ones, the Coulomb interactions in GaMo$_4$S$_8$ are expected to be weaker than in GaV$_4$S$_8$. Thus, in the following we will focus on the results obtained with $U_{eff}$=1 eV, if not otherwise explicitly noted. In the $U_{eff}$=1 eV case, the discrepancies with respect to experimental values for the lattice constants and rhombohedral angle are small (~-0.45% and ~+0.47% respectively), indicating the reliability of our DFT results.

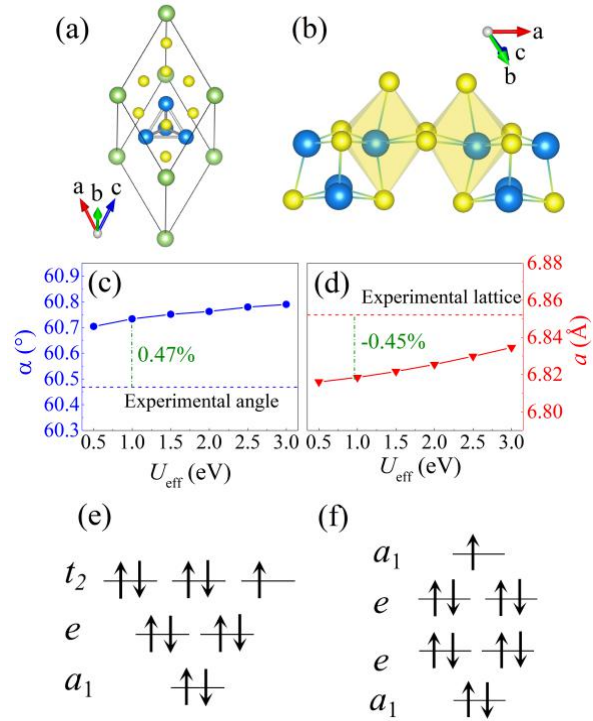

**Figure 1.** Structure of GaMo$_4$S$_8$. (a) The low symmetric structure (space group: $R3m$; point group $C_{3v}$). Green: Ga; Blue: Mo; Yellow: S. The distorted Mo$_4$ tetrahedron is also shown. (b) Cubane (Mo$_4$S$_4$)$^{5+}$ building blocks, highlighting the octahedral coordination of Mo with nearest-neighbour S atoms. (c-d) Optimized lattice constant and angle in the rhombohedral notation as a function of $U_{eff}$. The experimental values were measured at 8 K [38]. (e-f) The scheme of Mo$_4$ tetrahedron molecular orbitals arising from symmetry-allowed combinations of atomic $t_{2g}$ orbitals of octahedrally coordinated Mo atoms, with total 11 electrons, as in (e) the high symmetry strucure and (f) the low symmetry structure.

The isostructural (but not isoelectronic) GaV$_4$S$_8$ has been reported to display ferroelectricity in its low-temperature structure [12,14], which was later ascribed to a cooperative Jahn-Teller effect, removing an electronic degeneracy by locally elongating V$_4$ tetrahedra along the [111] direction [21]. Since the parent (high-temperature) structure belongs to the noncentrosymmetric $T_d$ point group, the Jahn-Teller distortion carries along an electric polarization, [21,39] which in GaV$_4$S$_8$ points towards the apical V$_1$ atom. The same mechanism is at play also in GaMo$_4$S$_8$, as sketched in Figs. 2(a-c), the only difference being that the Jahn-Teller effect induces a compression [Fig. 2(c)], rather than an elongation [Fig. 2(b)], of Mo$_4$ tetrahedra along the [111] direction. As discussed in Refs. [15,21] and in the next section, the opposite Jahn-Teller distortions can be rationalized by taking into account the different number of valence electrons in Mo and V lacunar spinels, leading to different partial filling of the tetrahedral molecular orbitals. As a consequence, the electric polarization in GaMo$_4$S$_8$ points in the opposite direction as compared to GaV$_4$S$_8$, namely from the apical Mo$_1$ atom to the tetrahedron base, as sketched in Fig. 2(c),

The ferroelectric polarization of GaMo$_4$S$_8$ is calculated using the Berry phase method, as shown in Fig. 2(d). At variance with what happens for lattice constants and rhombohedral angles, the polarization shows a rather strong dependence on the values of $U_{\text{eff}}$, varying in the range 1.8-0.7 μC/cm$^2$. In any case, the existence of polarization is unambiguous and its magnitude is of the order of 1 μC/cm$^2$. For $U_{\text{eff}}$ =1 eV, the polarization is estimated as ~1.4 μC/cm$^2$ (~3.1 μC/cm$^2$) using the relaxed (experimental) structure. Thus the polarization seems to be quite sensitive to structural changes, the relaxed structure showing changes in lattice constants below 0.5% and in bond lengths below 1% with respect to the experimental ones. In addition, a very recent work reported a switchable pyroelectric polarization ~0.2-0.4 μC/cm$^2$ for GaMo$_4$S$_8$ polycrystalline samples [23]. Such a value is smaller than our computational estimate, but it becomes in reasonable agreement with our predictions when considering their polycrystalline factor (typically 10%-30% of the single crystalline one). Furthermore, we also calculated the polarization of GaV$_4$S$_8$ (~2.55 μC/cm$^2$ with $U_{\text{eff}}$ =2 eV), that is consistent with the value of 2.43 μC/cm$^2$ reported in Ref. [21]. The polarization given by Wannier centers are 1.25 μC/cm$^2$ for GaMo$_4$S$_8$ and 2.00 μC/cm$^2$ for GaV$_4$S$_8$, in good agreement with the Berry phase method.

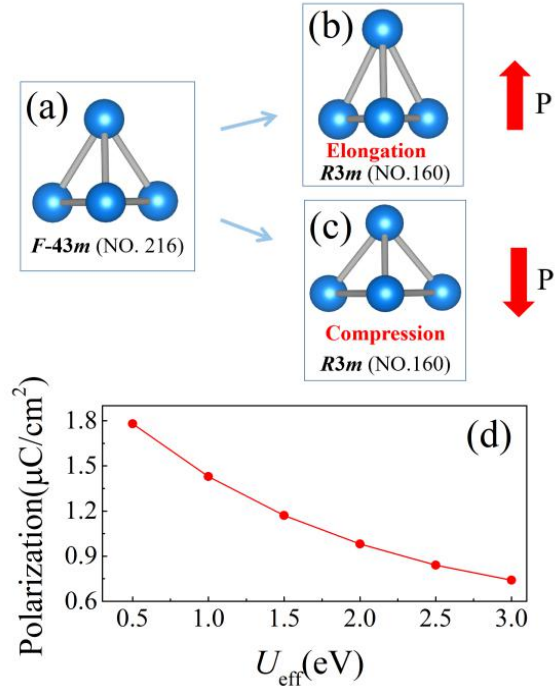

**Figure 2**. Polarization driven by Jahn-Teller distortion. Four Mo's form a tetrahedron. (a) Mo-tetrahedron in high symmetry (point group $T_d$) without Jahn-Teller distortion. (b-c) Mo-tetrahedron with Jahn-Teller distortions (elongated/compressed along the [111] direction). The directions of corresponding polarizations are indicated by arrows. In GaMo$_4$S$_8$, the tetrahedron is compressed. (d) Polarization as a function of $U_{eff}$.

*II.B Band structure and magnetism*

To further understand the molecular orbital character of Mo-tetrahedron, the projected band structures are plotted in Figs. 3(a-c). The bandwidths for bands around the Fermi level are quite narrow, since the electronic hopping between Mo-tetrahedra is weak. Thus, each Mo-tetrahedron is similar to a Mo$_4$ molecular unit. Without SOC, both the highest occupied molecular orbital (HOMO) and lowest unoccupied molecular orbital (LUMO) show the $a_1$ character. According to the density of states (DOS) (see Fig. 3(d)), there is a small Mott-Hubbard gap of ~0.21 eV, in agreement with previous predictions that gaps are within the range of 0.2~0.3 eV for $AM_4X_8$ family [40]. When including SOC (a required test due to the presence of heavy 4d element Mo), there is no big change with respect to the scalar relativistic band structure. This shows that SOC does not produce a severe rearrangement of the electronic levels.

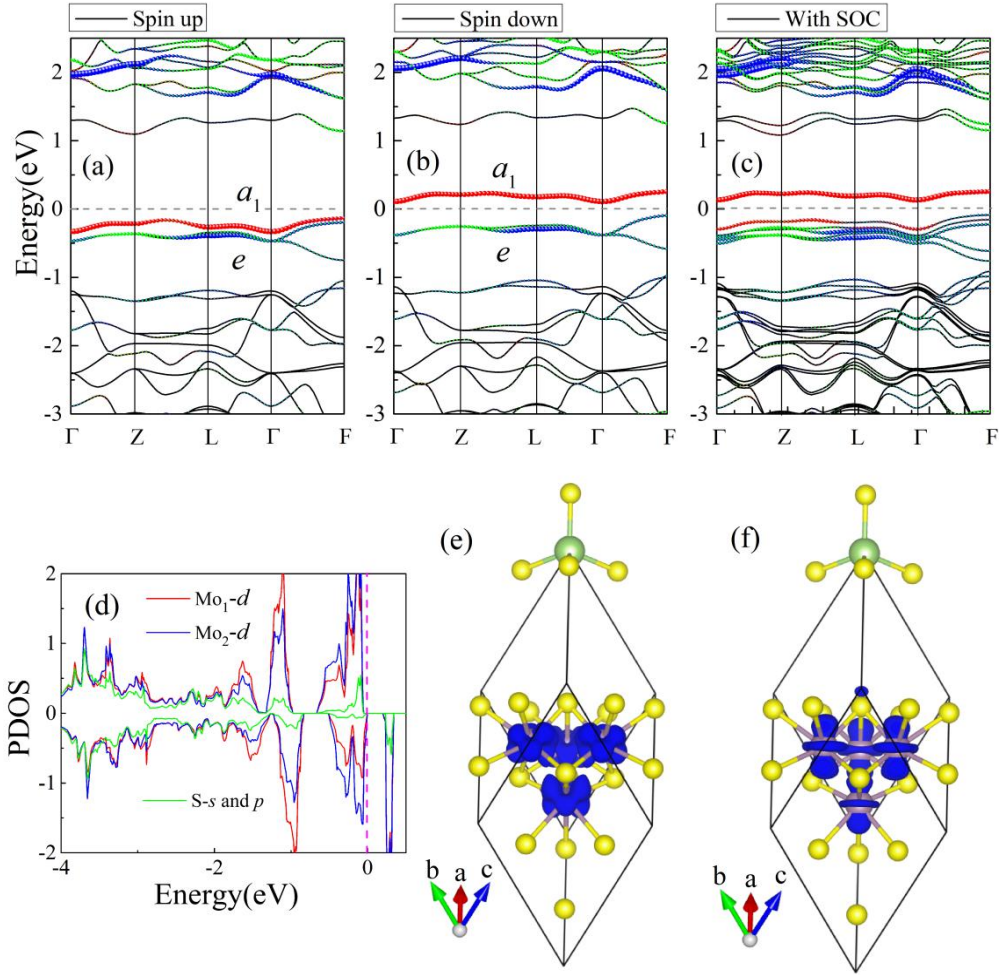

**Figure 3.** (a-b) Spin-up and spin-down electronic band structures of GaMo$_4$S$_8$. (c) Electronic band structure with SOC. (d) Atomic PDOS of an apical Mo$_1$ and a planar Mo$_2$, and a S with $s$ and $p$ orbitals. The pink dash line represents the Fermi energy. The shapes of (e) $e$ orbital and (f) $a_1$ orbital from partial density charge calculation.

The electronic structure can be understood from the molecular orbital viewpoint, as sketched in Figs. 1(e-f). Considering the values of Ga$^{3+}$ and S$^{2-}$, the average valence of Mo is +3.25. Thus the total $d$ electrons for each Mo$_4$ cluster are 11. In the high symmetric structure, the crystal field of each octahedrally coordinated Mo atom highlighted in Fig. 1(e) splits the atomic orbitals in higher-energy $e_g$ and lower-energy $t_{2g}$ orbitals in the local (octahedral) reference frame. Within each Mo$_4$ tetrahedron with $T_d$ symmetry, the atomic $t_{2g}$ states form molecular orbitals, with fully occupied $a_1$, $e$ states and partially occupied three-fold degenerate $t_2$ states, which are occupied by five electrons, as shown in Fig. 1(e). The partially filled degenerate electronic $t_2$ state strongly couples to a polar $T_2$ phonon mode, corresponding to a compression/elongation of the Mo$_4$ tetrahedral unit which lowers its symmetry from $T_d$ to $C_{3v}$, and splits into a doubly degenerate $e$ state and a single-degenerate $a_1$ state. In order to remove the electronic degeneracy,Within each Mo4 tetrahedron the $e$ states are pushed to lower energy and are fully occupied by four electrons, while the single-degenerate $a_1$ state is promoted to higher energy (occupied by one electron), as schematically depicted in Fig. 1(f) and consistently with the band structures displayed in Fig. 3(a)-(c) showing both HOMO and LUMO states with $a_1$ character. The very same mechanism has been discussed for GaV$_4$S$_8$ [21], where, however, the molecular orbitals are occupied by seven electrons, which leaves a single electron in the three-fold degenerate $t_2$ state. In order to remove the degeneracy, the Jahn-Teller distortion of V$_4$ tetrahedron promotes the $e$ states to higher energy, thus leaving the single electron in the $a_1$ state which is now pushed to lower energy, opposite to GaMo$_4$S$_8$. This different behavior can be rationalized taking into account that, in the rhombohedral reference frame, where the $z$ axis is parallel to the [111] polar direction, the $a_1$ state has a predominant $d_{z^2}$ character. As a consequence, the $a_1$ state can be pushed to higher (lower) energy by a compression (elongation) of the transition-metal tetrahedral unit along the $z$ axis. [15,21].

In both cases, the presence of an unpaired spin in the valence manifold results in a total magnetization of 1 $\mu_B$ per transition-metal tetrahedron.

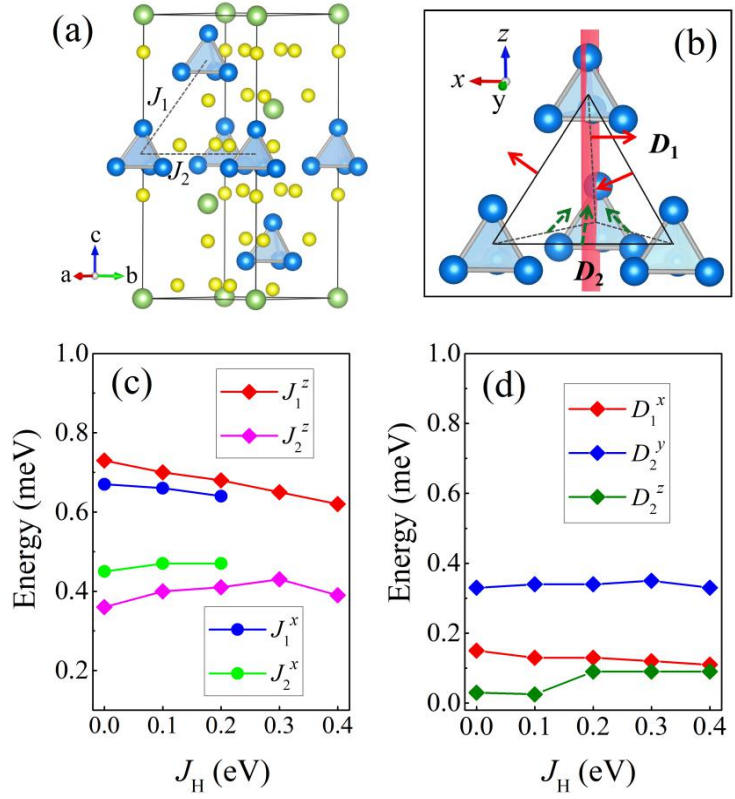

**Figure 4**. (a) Structure of GaMo$_4$S$_8$ in the hexagonal setting. The exchange interactions are marked as $J_1$ (inter-plane) and $J_2$ (intra-plane). (b) The DM vectors denoted as $D_1$ (inter-plane) in red and $D_2$ (intra-plane) in green. The vector directions are determined by the local symmetries: $D_1=(D_1^x, 0, 0)$; $D_2=(0, D_2^y, D_2^z)$. Note that the directions of $D_1$ and $D_2$ for each bond will rotate with the bond vector, according to crystal symmetries with triple rotational symmetry. (c-d) Magnetic coefficients as a function of $J_H$, while $U-J_H$ is fixed to 1 eV. (c) $J$'s with magnetic moments pointing to the $z$ axis and (d) the components of $D$'s.

According to Ref. [32], the magnetic moments in GaV$_4$S$_8$ are mainly located on the apical V$_1$ ion, while the planar V$_2$ ions only contribute a little spin density. This disproportionation of magnetic moments usually occurs when accompanied by charge ordering, driven by the strong Hubbard repulsion. The situation is quite different in GaMo$_4$S$_8$. Our DFT calculation suggests that the magnetic moments are in the range 0.21-0.25 $\mu_B$ for Mo$_1$ and all Mo$_2$'s, implying quite homogeneous spin density within the Mo-tetrahedron. The atomic projected DOS (PDOS) also shows the very similar result for Mo$_1$ and Mo$_2$ (see Figs. 3(d)). Furthermore, the spatial distributions of partial density charge also confirm the homogeneous $a_1$ and $e$ molecular orbitals (Figs. 3(e-f)). The distinct character of GaMo$_4$S$_8$ can be attributed to the more spatial-extended $4d$ orbitals of Mo, which is beneficial to form orbital hybridization within tetrahedron.

Based on the DFT results, a classical spin Heisenberg model is constructed to model the magnetic properties of GaMo$_4$S$_8$. Taking each Mo-tetrahedron as a magnetic unit, there are two types of bonds between different tetrahedra, with distances 6.82 Å out-of-plane and 6.90

Å in-plane. As shown in Fig. 4(a), the out-of-plane nearest exchange is marked as $J_1$, while the in-plane nearest exchange is marked as $J_2$. Similar notation is also applied to the DM vectors, $D_1$ and $D_2$. Since the next-nearest neighbour distances are much larger (with 9.70 Å out-of-plane and 11.95 Å in-plane), the corresponding exchange constants are negligible. Then the model Hamiltonian can be written as:

$$H = H_{ex} + H_{DM} + H_{SIA} + H_{Zeeman}$$

$$H_{ex} = -J_{1x}\sum_{<i,j>}(S_i^x \cdot S_j^x + S_i^y \cdot S_j^y) - J_{2x}\sum_{<i,j>}(S_i^x \cdot S_j^x + S_i^y \cdot S_j^y) - J_{1z}\sum_{<i,j>}S_i^z \cdot S_j^z - J_{2z}\sum_{<i,j>}S_i^z \cdot S_j^z$$

$$H_{DM} = -D_1 \cdot \sum_{<i,j>}S_i \times S_j - D_2 \cdot \sum_{<i,j>}S_i \times S_j$$

$$H_{SIA} = -A\sum_i (S_i^z)^2$$

$$H_{Zeeman} = -h \cdot \sum_{<i>} S_i$$

where $S_i$ represents the Heisenberg spin with unit length on site $i$. $H_{ex}$ is the symmetric exchange interaction, including a XXZ-type anisotropy where $J_{1x}$ and $J_{2x}$ represent the spins pointing in-plane while $J_{1z}$ and $J_{2z}$ represent spins pointing out of plane. $H_{DM}$ includes the DM interactions both in-plane and out of plane. $H_{SIA}$ is the single-ion anisotropy with coupling constant $A$. $H_{zeeman}$ is the Zeeman energy due to magnetic field $h$. It should be noted that our model differs from the Heisenberg XXZ model previously studied for GaV$_4$S$_8$ [13,14], which was defined on a two-dimensional triangular lattice (thus neglecting $J_1$ and $D_1$ interactions) and did not include the single-ion anisotropy.

In order to estimate the magnetic parameters entering in the model Hamiltonian, we calculated the total energies, including SOC, of several collinear and non-collinear spin configurations (cfr Ref. [32]). The model parameters were then extracted by mapping the DFT total energy to the model shown above. Since the magnetic anisotropic energy (MAE) depends on the choice of Hund coupling term $J_H$ [41], the Liechtenstein approach [31] was adopted to apply the Hubbard correction, keeping $U$-$J_H$=1 eV (the simplified Dudarev approach corresponds to the $J_H$=0 eV limit.). The estimated exchange-interaction parameters are shown in Fig. 4(c-d), displaying only slight changes as a function of $J_H$. Both inter-plane and intra-plane symmetric exchange-interactions $J_1$ and $J_2$ are ferromagnetic, the former being roughly twice the latter. The estimated anisotropy of the dominant symmetric exchange is $J_{1z}/J_{1x}$ = 1.07. On the contrary, and analogously to GaV$_4$S$_8$ [32], the intra-plane DM interaction $D_2$ is significantly larger than inter-plane $D_1$. As expected, here the absolute value of DM vectors are larger than those in GaV$_4$S$_8$ [32], due to the stronger SOC of 4$d$ orbitals.

**Table 1.** The anisotropic symmetric exchange coefficient and the single-ion anisotropic (SIA) coefficient A used in MC simulations.

|  | $J_{1x}$ (meV) | $J_{2x}$ (meV) | $J_{1z}$ (meV) | $J_{2z}$ (meV) | $A$ (meV) |
|---|---|---|---|---|---|
| $J_H$=0 eV | 0.67 | 0.45 | 0.72 | 0.36 | 0.17 |
| $J_H$=0.1 eV | 0.66 | 0.47 | 0.70 | 0.40 | 0.32 |

*III.C Magnetic phase transitions*

Based on the Heisenberg spin model and using the aforementioned DFT coefficients, MC simulations are performed to study the magnetism under magnetic field at finite temperatures. Starting from $J_H$=0, the magnetic ordering is simulated as a function of temperature. As shown in Fig. 5(a), the heat capacity peaks at ~17.5 K, very close to the experimentally reported temperature, 19.5 K, although the magnetization remains small in the whole temperature range. A careful analysis on spin structure factor as well as MC snapshot confirms the non-collinear spiral texture in the *ab* plane of hexagonal lattice as the ground state (see Fig. 6(c)). The wave length of a spiral cycloid is about 21 u.c. (~14.6 nm).

By applying a magnetic field along the *c*-axis (i.e. perpendicular to the spiral plane), the spiral texture transforms to the Néel-type skyrmion lattice (see Fig. 6(d) for example) and finally to a full ferromagnetic state, provided the magnetic field is large enough. The global phase diagram for $J_H$=0 eV is shown in Fig. 6(a). The typical spin textures of both cycloid and skyrmion phases are also plotted in Fig. 6(c-d), with their corresponding in-plane spin structure factor. In the skyrmion phase of $J_H$=0 eV, a typical radius ($r_{Mo}$) of each skyrmion is estimated to be 6.9 nm, much smaller than most previously reported skyrmions (including the sister compound GaV$_4$S$_8$: $r_V$~11 nm [14]). To our knowledge, only the hexagonal Fe film of one-atomic-layer thickness on the Ir(111) surface shows a smaller size ($r_{Ir}$~1 nm) Néel-type skyrmions so far [42].

The reason for such a small size skyrmion, i.e. high dense skyrmion lattice, might be the large $|D|/J$ ratio, due to the large SOC of 4d orbitals. In fact, in our calculation $D_2/J_{2x}$ (in-plane) is 0.74, while $D/J_\perp$ of GaV$_4$S$_8$ ($D$ is the in-plane DM vector, and $J\perp$ represents the in-plane exchange interaction) in Ref. [14] is only 0.35 which is about half of that in GaMo$_4$S$_8$. As we known, the size of skyrmion lattice is determined by the competition of DM interaction and exchange $J$, and in inverse ratio to the the value of $D/J$. When comparing our MC results with $r_{Mo}$ in Ref. [14], the ratio of $r_{Mo}/r_V$ is 0.64, which is semiquantitatively agree with (but slight larger than) the expectation from $D/J$. The slight quantitative bias is also reasonable considering the enhanced anisotropy energy (the single-ion *A* and anisotropic exchanges especially for $J_2$) in GaMo$_4$S$_8$. Because the anisotropy will suppress the noncollinear texture as well as skyrmion. Also in Ref. [14], the model is purely on a two-dimensional lattice, without $A$ and $J_2$. Our model lattice is three-dimensional, with more realistic interactions considered.

When increasing the values of $J_H$, although the exchange interactions $J$'s and DM vectors don't change much, the SIA sizeably increased with $J_H$ =0.1 eV comparing to $J_H$ =0 eV, as shown in Table 1. We also plot the phase diagram (shown in Fig. 6(b)) with the parameters calculated in the case of $J_H$ =0.1 eV. One therefore observes that the skyrmion phase still appears in a small area, and also in the high temperature range.

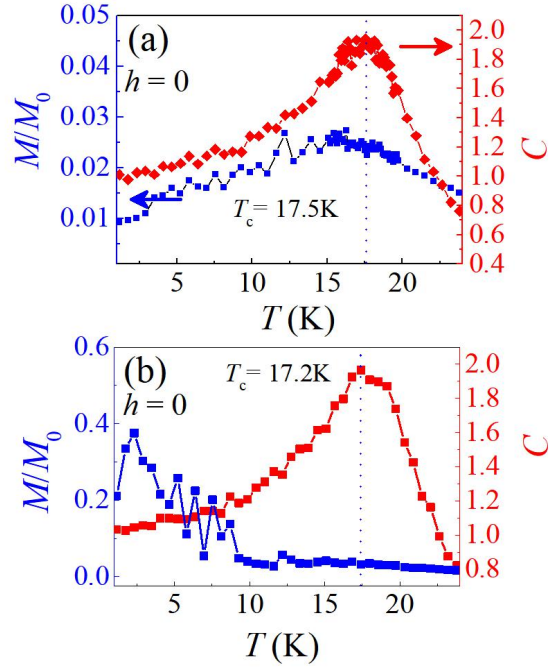

**Figure 5**. MC results as a function of temperature for (a) $J_H$=0 and (b) $J_H$=0.1 eV. The low magnetization ($M$) and peak of heat capacity (right axis) suggests non-ferromagnetic (ferromagnetic) ground state for $J_H$=0 ($J_H$=0.1 eV). $h$ denotes the external magnetization field and $M_0$ denotes the saturated magnetization.

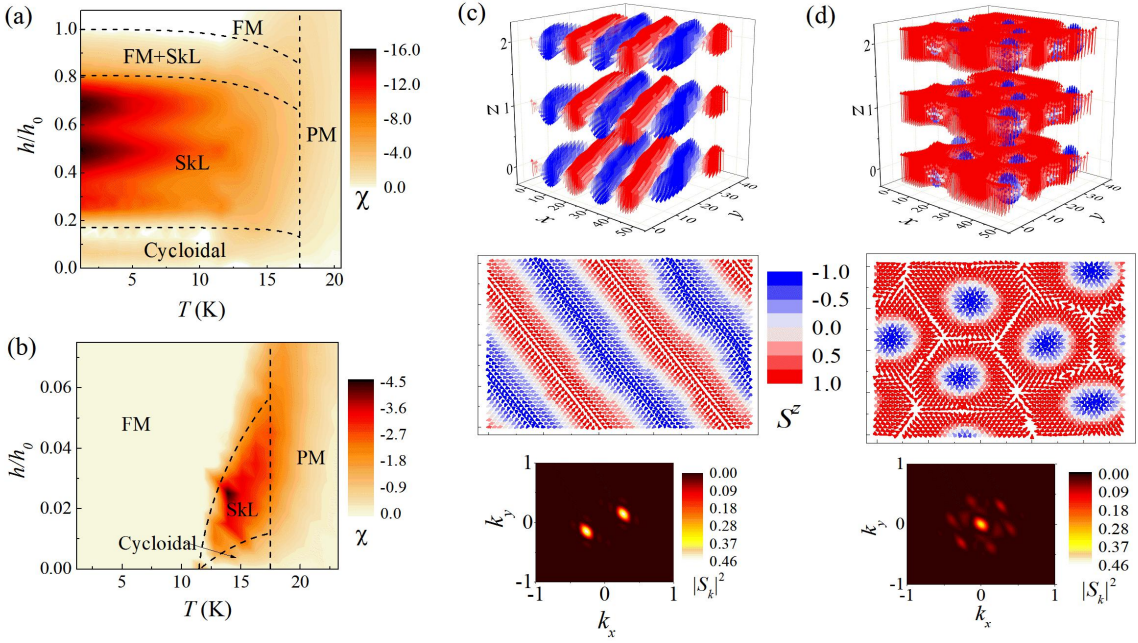

**Figure 6**. MC phase diagrams for (a) $J_H$=0 eV and (b) $J_H$=0.1 eV. The phase boundaries are determined by heat capacity, magnetic susceptibility, spin structure factors, local spin chirality, as well as snapshots. The peaks of capacity and susceptibility mark the transition points, while the magnetization and local chirality can distinguish different phases, which are further verified by real space snapshots as well as spin structure factors. All these three quantities lead to consistent results. (c) Upper: three-dimensional snapshot of cycloid spin order. Middle: in-plane spin structure factor. Lower: in-plane spin structure factors. (d) Same as (c) but for the skyrmion lattice. The bright points in spin structure factors denote the modulation wave vectors. The double peaks correspond to cycloid spin order while the six-fold peaks reflect the skyrmion lattice. $h_0$ denotes the saturation magnetization at low temperature.

*III.D Comparison with GaV$_4$S$_8$*

Finally, it is worth to highlight the differences between GaMo$_4$S$_8$ and GaV$_4$S$_8$, although they are quite similar in structure. First, the different occupation of electrons in *d* orbitals leads to an opposite sign of polarization upon Jahn-Teller distortions, as found in Sec. III.A and discussed in Sec. III.B. Second, due to the more-extended 4*d* orbitals, the magnetic disproportionation in V-tetrahedrons reported in GaV$_4$S$_8$ [32] is not observed in Mo-tetrahedrons of GaMo$_4$S$_8$. Third, thanks to the larger SOC effect on 4*d* orbitals, the DM interaction in GaMo$_4$S$_8$ is stronger than in GaV$_4$S$_8$. These enhanced DM interactions can lead to non-collinear spin textures with larger spin angles between nearest neighbors. Thus, the corresponding period of cycloid and size of skyrmions are much smaller in GaMo$_4$S$_8$, as verified in our MC simulations. However, also due to the larger SOC effect, the MAE may be also larger, which can help the ferromagnetic exchange to stabilize the collinear ferromagnetic ordering, rather than non-collinear ones. For example, the ground state calculated for $J_H$=0.1 eV is ferromagnetic and the skyrmions can only exist in a narrow temperature region just below the Curie temperature. In this sense, the strong SOC is a double-edged sword in pursuing the stabilization of the skyrmion phase in $AM_4X_8$ spinels.

Although the MAE, resulting from the exchange anisotropy and the SIA, depends to a certain degree on the choice of $J_H$ in DFT calculations, the skyrmion phase can appear with reasonable ASE and SIA values. Therefore, our results suggest that $GaMo_4S_8$ can host high-density skyrmions under magnetic field. However, one should keep in mind that, in case the SIA is experimentally found to be very strong, the ground state could then be ferromagnetic. Even in this case, skyrmions could still be observed with the help of thermal activation, similar to what happens in $GaV_4S_8$. Nonetheless, both the predicted phases, i.e. ferroelectric-skyrmion or ferroelectric-ferromagnet, are interesting, the results being valuable and undoubtedly calling for further experimental studies.

## IV. Conclusion

The structural and electronic properties, electric polarization, as well as magnetic interactions in $GaMo_4S_8$ have been systematically studied via DFT simulations. The exchange coupling constants and DM interactions show comparable values, their competition inducing very peculiar magnetic properties. Non-collinear magnetic states were predicted by using the MC approach with model parameters estimated from DFT. Importantly, the skyrmion lattice phase was predicted to probably appear in a large area of the phase diagram, which depends on the intensity of Hund coupling. Despite the similarity between $GaV_4S_8$ and $GaMo_4S_8$, the peculiarities of 4$d$ orbitals (i.e. larger spatial extension as well as stronger SOC) lead to different results, in terms of increased phase stability for the skyrmionic phase and appealing reduced size of the skyrmions. Our predictions are therefore expected to stimulate future experiments, aiming at studying high-density skyrmion lattices in lacunar spinels and related systems.

**Acknowledgment**

S.D., H.M.Z, and J.C. were supported by National Natural Science Foundation of China (Grant No. 11834002 and 11674055). H.M.Z. was partially supported by the China Scholarship Council. K.Y. was supported by JSPS Kakenhi Grant Number 18H04227 and by JST CREST Grant Number JPMJCR18T1, Japan. The work at CNR-SPIN was performed within the framework of the Trieste Nanoscience Foundry and Fine Analysis (NFFA-MIUR Italy) project. We thank the Tianhe-II of National Supercomputer Center in Guangzhou (NSCC-GZ) and Big Data Center of Southeast University for providing the facility support on the numerical calculations.